# Regulating effect of biaxial strain on electronic, optical and photocatalytic properties in promising $X_2$PAs ($X$ = Si, Ge and Sn) monolayers

*Qi-Wen He [a], Yang Wu [a], Chun-Hua Yang [a], He-Na Zhang [a], Dai-Song Tang [a], Cailong Liu [b], and Xiao-Chun Wang [a]\**

[a] Institute of Atomic and Molecular Physics, Jilin University, Changchun 130012, China

[b] School of Physics Science and Information Technology, Liaocheng University, Liaocheng 252000, China

**ABSTRACT:**

Photocatalytic water splitting is an effective way to obtain renewable clean energy. The challenge is to design tunable photocatalyst to meet the needs in different environments. At the same time, the oxygen and hydrogen evolution reactions (OER and HER) on the photocatalyst should be separated, which will be conducive to the separation of products.

---

\* Corresponding author *E*-mail: wangxiaochun@tsinghua.org.cn (Xiao-Chun Wang)

The electronic, optical and photocatalytic properties of Janus $X_2$PAs ($X$ = Si, Ge and Sn) monolayers are explored by first-principles calculation. All the strain-free $X_2$PAs monolayers exhibit excellent photocatalytic properties with suitable band edge positions straddling the standard redox potential of water and large visible light absorption coefficients (up to $10^5$ cm$^{-1}$). Interestingly, the intrinsic internal electric field is favorable for separating photogenerated carriers to different surfaces of the monolayer. It contributes to realize the OER and HER separated on different sides of the monolayer. In particular, the energy band edge positions of $X_2$PAs monolayers can be well adjusted by biaxial strain. Then it can effectively modulate photocatalytic reactions, suggesting $X_2$PAs monolayers can be a piezo-photocatalytic switch between the OER, HER and full-reaction of redox for water splitting. This investigation not only highlights that the photocatalyst $X_2$PAs monolayers with the separated OER and HER can be effectively tuned by the mechanical strain, but also provides a new strategy for designing highly adaptable and tunable piezo-photocatalysts.

1. Introduction

The development of clean and renewable hydrogen energy is of fundamental importance for coping with global energy shortage and environment deterioration. A lot of research has been carried out since photocatalytic water splitting for hydrogen evolution was first proposed in 1972 [1]. In earlier studies, effective photocatalysts for water splitting were concentrated on three-dimensional (3D) materials, such as metal nitrites, oxides, and sulfides [2-4]. However, 3D photocatalysts with wide band gap and low carrier mobility exhibit low efficiency of solar energy conversion to hydrogen production, which hinders their further application in photocatalysis field [5, 6]. In contrast, two-dimensional (2D) materials with ultrathin nature can shorten the distance of carrier migration, thus reducing photogenerated carriers recombination. So far, 2D materials have attracted a growing interest for their outstanding photocatalytic properties, such as graphene, $AgBiP_2Se_6$, $MoSi_2N_4$, nitrides and oxides [7-14].

2D materials with Janus geometry have gotten much attention for the new properties arising from the broken out-of-plane symmetry [15-22]. More interesting, Yang's group have proposed a new photocatalytic model with an intrinsic dipole that can relive the restriction on the band gap requirement of photocatalysts [23]. The potential applications of piezoelectricity in photocatalysis are revealed. For example, 2D Janus $M_2X_3$ ($M$ = Al, Ga, In; $X$ = S, Se, Te) have formed a series of photocatalytic materials, which overcame the band-gap limitation of water decomposition and the traditional theoretical efficiency (18%) limitation [24]. Nevertheless, the search for the tunable high-performance photocatalysts is still ongoing. Recently, it has been reported that biaxial strain can adjust the band gaps, band edge positions and optical properties of group-III monochalcogenides ($MX$, $M$ = Ga or In, $X$ = S, Se, or Te), thus improving the solar energy conversion efficiency [25]. The Janus $X_2$PAs ($X$ = Si, Ge and Sn) monolayers with in-plane and out-of-plane piezoelectricity are dynamically and thermally stable [15].

Motivated by this exciting strategy, in this work, we focus on the regulating effect of biaxial strain on electronic, optical and photocatalytic properties in Janus $X_2$PAs ($X$ = Si, Ge and Sn) monolayers based on first-principles calculation. Firstly, the Bader charge analysis is performed to explore the internal charge distribution in $X_2$PAs monolayers. Next, the electronic properties of strain-free $X_2$PAs are studied by HSE06 method. Then, the electronic properties of $X_2$PAs under biaxial strain are calculated to investigate how the biaxial strain can tune the electronic and photocatalytic properties. Finally, the optical properties of $X_2$PAs with and without strain are discussed. These results disclose the fascinating properties of 2D Janus $X_2$PAs monolayers, effectively demonstrating their potential as the switch between the OER, HER and full-reaction for water splitting.

## 2. Methods

The Projector Augmented Wave (PAW) method of Vienna *ab initio* simulation package (VASP) is applied to perform all the first-principles calculations in this work [26-28]. In the calculation, the Perdew-Burke-Ernzerhof (PBE) electron exchange correlation functional of the generalized gradient approximation (GGA) is adopted [29]. The electronic and optical properties are evaluated by Heyd−Scuseria−Ernzerhof (HSE06) hybrid functional [30]. Spin polarization and dipole moment are taken into account to accurately calculate the results. In order to avoid interlayer interaction, the thickness of vacuum layer along the z direction is more than 25 Å. The energy convergence criterion for structural optimization is set to $10^{-6}$ eV, while the force converges to less than $10^{-4}$ eV/Å on each atom. A 7 × 7 × 1 Monkhorst-Pack [31] k-point grid is used to sample the first Brillouin area with the plane wave cutoff energy set to 500 eV. The bonding characteristics of atoms are analyzed through Bader method [32].

## 3. Result and discussion

### 3.1 Geometrical structure and electronic properties of Janus $X_2$PAs monolayers

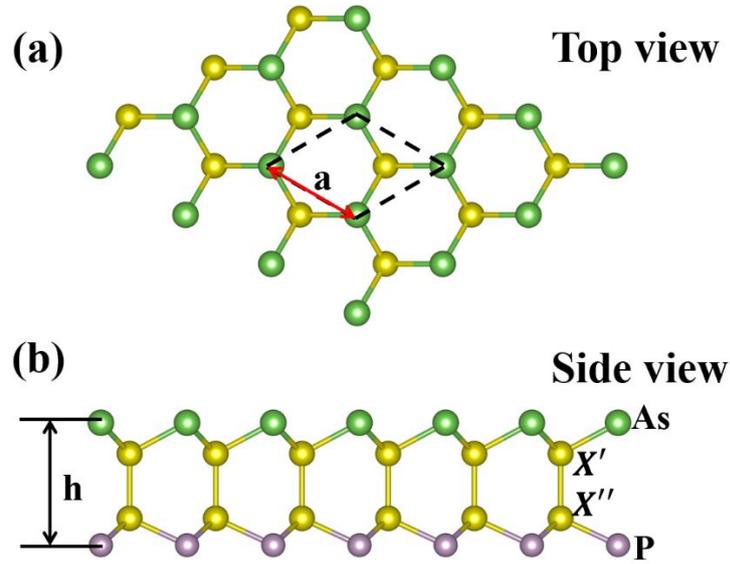

**Fig. 1.** (a)Top and (b)side views of the optimal structure for Janus $X_2$PAs monolayers. *h* is the thickness of the monolayer and *a* is lattice constant. The primitive cell marked by black dotted lines.

As shown in Fig. 1, each $X_2$PAs (X = Si, Ge and Sn) monolayer has hexagonal Janus structure, and the stacking sequence of atomic layers is As-$X'$-$X''$-P ($X'$, $X''$ = Si, Ge and Sn). The $X_2$PAs monolayers with both of in-plane and out-of-plane asymmetry may lead to the charge transfer in different atom layers. To explore the charge distribution in $X_2$PAs monolayers, Bader charge analysis is calculated, summarized in Table 1. The values display the change in charge of the atom relative to its initial charge. The negative charges are mainly concentrated in P and As atoms. Since P atom is more electronegative than As atom, P atom gather more negative charges. Hence, the charge transfer between P-$X''$ bond and As-$X'$ bond is unbalance in the $X_2$PAs monolayers, with a positive polarization charge in the lower side (P-$X''$ region) and a negative polarization charge in the upper surface (As-$X'$ region). The charge redistribution generates a total dipole moment with the direction from P-$X''$ bond to As-$X'$ bond, giving rise to an internal electric field pointing from the lower surface side of the monolayer to the upper surface side. The internal electric field, as a direct driving force, plays a crucial role in propelling the separation of photogenerated carrier, and preventing photogenerated carrier recombination.

**Table 1**

The value (unit in |e|) of charge transfer between each atom in $X_2$PAs monolayers based on Bader charge analysis.

| Material | As | $X'$ | $X''$ | P |
|---|---|---|---|---|
| Si$_2$PAs | -0.73 | 0.68 | 1.20 | -1.14 |
| Ge$_2$PAs | -0.28 | 0.25 | 0.50 | -0.46 |
| Sn$_2$PAs | -0.45 | 0.42 | 0.67 | -0.62 |

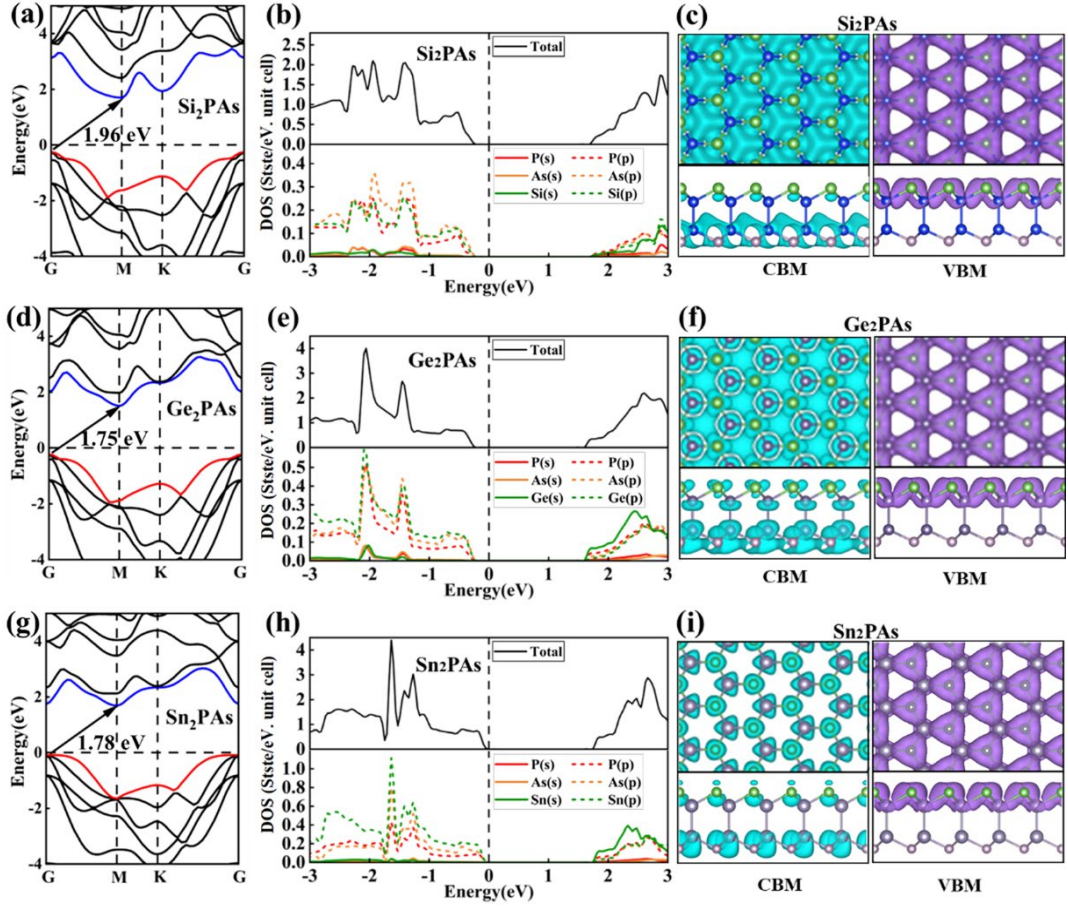

**Fig. 2.** (a), (d), (g) Band structure of $X_2$PAs monolayers based on the HSE06 functional. (b), (e), (h) TDOS and PDOS of each atom in the $X_2PAs$ monolayers. (c), (f), (i) Top and side views of the charge densities at the CBM (blue) and VBM (purple). The value of the isosurface is 0.013 e/Å$^3$.

Attaining accurate band structures is a fundamental step in characterizing the electronic properties. Hence, all the band structures calculation are based on the precise HSE06 method. As displayed in Fig. 2 (a, d, g), all $X_2$PAs materials exhibit indirect band gap. The conduction band minimum (CBM) of $X_2$PAs is located at M point, while the valence band maximum (VBM) of $X_2$PAs is located at G point. The largest band gap is 1.96 eV in $Si_2PAs$. The smallest band gap is 1.75 eV in $Ge_2PAs$. These are in agreement with the value reported by the previous results [33]. Notably, the band gaps of all $X_2PAs$ monolayers are larger than 1.23 eV that is the minimum Gibbs free energy required for water splitting, suggesting their potential photocatalytic application.

In Fig. 2 (b, e, h), it is obviously that the CBM and VBM are contributed by different atoms. The s orbital of $X$ ($X$ = Si, Ge and Sn) and p orbitals of P make major contribution to CBM, while the VBM is mainly attributed to the p orbitals of $X$ and p orbitals of As. These are consistent with previous reports [34]. The spatial distribution of VBM and CBM are shown in Fig. 2 (c, f, i). The charge density of CBM is mainly distributed in the P-$X''$ bond region, with a small amount of charge at the As-$X'$ bond region, whereas that of VBM is only distributed in P-$X''$ bond region. Hence, the oxygen evolution reactions (OER) may take place on upper surface (As-$X'$ region), while the hydrogen evolution reactions (HER) mainly at lower side (P-$X''$ region), which lead to a good spatial separation between photogenerated electrons and holes. The unique properties could reduce the possibility of electron-hole pair recombination, thus ensuring the high efficiency of photocatalytic reaction.

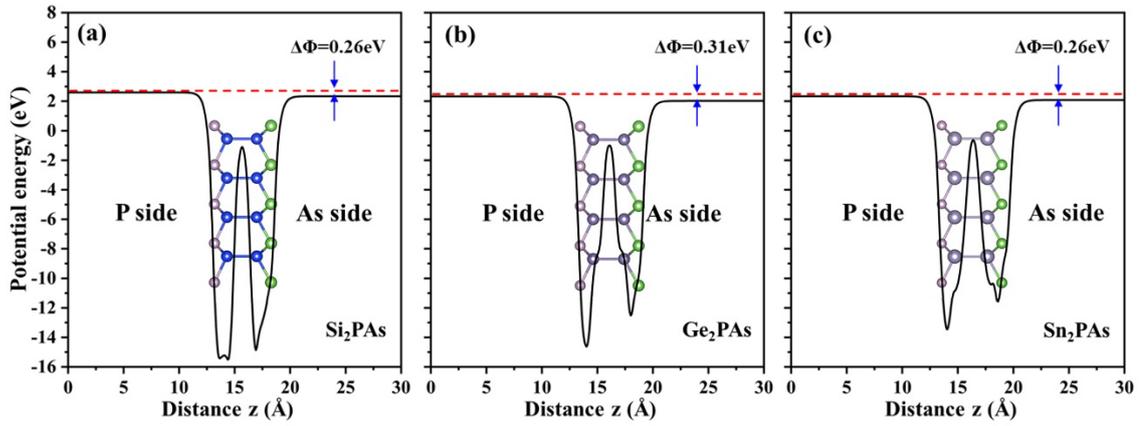

**Fig. 3.** The electrostatic potential difference of (a) Si$_2$PAs, (b) Ge$_2$PAs and (c) Sn$_2$PAs along the z direction.

The electrostatic potential differences of $X_2$PAs ($X$ = Si, Ge and Sn) along z direction are presented in Fig. 3. On account of the vertical asymmetry of Janus structure, there is an electrostatic potential difference between the upper and lower surface of $X_2$PAs monolayer within the range from 0.26 to 0.31eV. According to Bader charge analysis, the negative charge of P atom in $X_2$PAs monolayers is larger than that of As atom, therefore the electrostatic potential at P atom is lower than that at As atom. It is worth noting that the electrostatic potential of the vacuum layer on both sides of the monolayer is the result of the combined action of all four atom-layers. The P side is close to the P-$X''$ bond region with positive polarization charge, while the As side is close to the As-$X'$ bond region with negative polarization charge. Hence, the electrostatic potential of the P side is higher than that of the As side. The P side with a high electrostatic potential is conducive to attracting photogenerated electrons to P-$X''$ bond region, thus facilitating the HER process on the P-$X''$ bond region of the monolayer. On the contrary, the As side with a low electrostatic potential will attract the photogenerated holes to As-$X'$ bond region, thus impelling the OER process on the As-$X'$ bond region of the monolayer. Owing to the internal electrostatic potential difference, $X_2$PAs monolayers have a broad application prospect in low-cost photocatalytic water decomposition to hydrogen production.

**3.2 Tuning band edge of $X_2$PAs monolayers by biaxial strain**

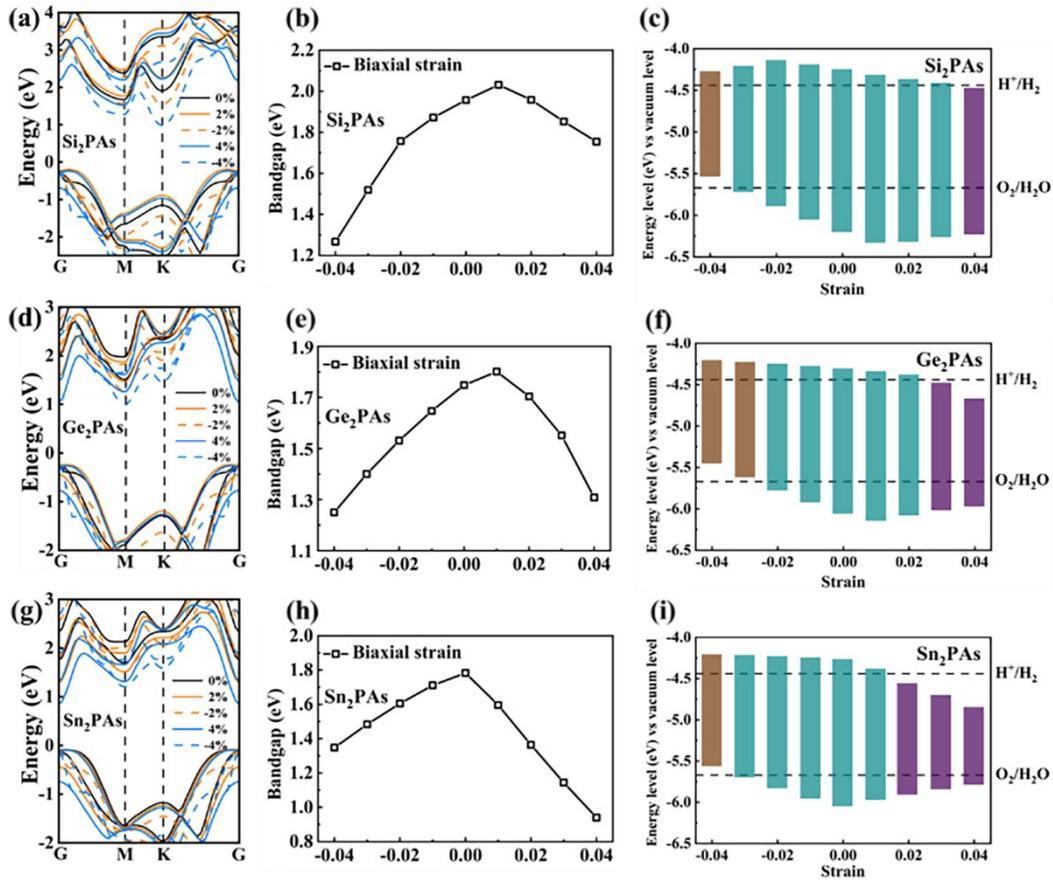

**Fig. 4.** (a),(d),(g) Band structures of $X_2$PAs monolayers.(b),(e),(h) Bandgap variation with in-plane biaxial strain based on HSE06. (c),(f),(i) Strain effects on band edge positions of $X_2$PAs monolayers in reference to the vacuum level.

Strain engineering has been proved to be an effective method for adjusting electronic and optical properties both of experimentally and theoretically [35, 36]. To research the effect of mechanical strain on band structures of $X_2$PAs ($X$ = Si, Ge and Sn) monolayers, biaxial strains within the range of −4% to +4% are applied to these monolayers.

Fig. 4 (a, d, g) show the VBM (CBM) positions of $X_2$PAs monolayers as a function of biaxial strain. Under the strain in the range from -4% to +4%, all the $X_2$PAs monolayers are keeping the indirect bandgap property, while the positions of VBM and CBM are partially changed. As tensile strain increases, the VBM of $Si_2$PAs and $Ge_2$PAs have a tendency to move from G point to K point. However, the CBM of $Ge_2$PAs and $Sn_2$PAs have a tendency to move from M point to G point. As the absolute value of compressive strain increases, the CBM of $Si_2$PAs at -2% strain changes from M point to K point, while the positions of CBM (VBM) of $Ge_2$PAs and $Sn_2$PAs remain unchanged. The band gaps of $X_2$PAs monolayers are further analyzed (Fig. 4 (b, e, h)). As analyzed, the band gap of $Sn_2$PAs with smaller Young's modulus than $Si_2$PAs and $Ge_2$PAs changes with the increase of biaxial strain in more obvious linear in two parts. As compressive strain increases, the band gaps of $X_2$PAs monolayers decreases monotonically. When the compressive strain reaches -4%, the band gap width of $Si_2$PAs is narrowed by 0.69 eV, which is the largest reduction among the three materials. Furthermore, the band gaps of $Si_2$PAs and $Ge_2$PAs both reach their maximum width with the tensile strain of +1%, then decrease as further tensile strain increases. Whereas, the band gap of $Sn_2$PAs decreases under both of tensile and compressive strains. Under the tensile strain of +4%, the band gap of $Sn_2$PAs decreases by 0.84 eV, which is 1.94 times of the band gap reduction caused by compressive strain.

With a view to further explore the effect of strain on the photocatalytic properties of $X_2$PAs monolayers, the biaxial strain effect on the band edge positions with respect to the vacuum level is analyzed. As shown in Fig. 4 (c, f, i), the band edge position changes with the increase of biaxial strain is not monotonic. This is mainly in consequence of the variation of CBM and VBM positions of $X_2$PAs monolayers. Intriguingly, the variation trend of VBM is consistent with that of Fermi level, while the biaxial strain increases. For example, the lowest Fermi level of $Sn_2$PAs is at 0% strain. Similarly, the lowest position of VBM in Fig. 4 (i) is also at 0% strain. $Si_2$PAs and $Ge_2$PAs exhibit the similar phenomenon. It is worth pointing out that the photocatalytic abilities of $X_2$PAs monolayers can be well tuned by macroscopic strain. For example, under the biaxial strain ranging from -3% to +1%, the band edge positions of $Sn_2$PAs can match the redox reaction requirement of photocatalytic water splitting, indicating its photocatalytic potential for both HER and OER. When the compressive strain is less than -4%, the CBM and VBM of $Sn_2$PAs are above the hydrogen reduction level (−4.44 eV at pH = 0) and the oxidation level of oxygen (−5.67 eV at pH = 0), respectively. It indicates that $Sn_2$PAs shows high reducibility, and is only favorable for the hydrogen evolution reactions (HER). As the tensile strain is more than +2%, the VBM and CBM of $Sn_2$PAs are under the hydrogen reduction level and the oxygen oxidation level, respectively. It exhibits only high oxidation for the oxygen evolution reactions (OER). The similar phenomenon is shown in the other two materials. Combined with elastic property analysis, the band edges positions of $X_2$PAs with smaller Young's modulus are easier to be moderated by strain. Specifically, the band edge positions of Sn$_2$PAs with the smallest Young's modulus in $X_2$PAs monolayers are easiest to be tuned by strain, so it is more suitable to be used as a switch between the OER, HER and full-reaction for water decomposition. While the band edges positions of Si$_2$PAs with the largest Young's modulus in $X_2$PAs monolayers is not sensitive to strain, thus it is more appropriate for photocatalysts in environments where the full-reaction of water splitting is only required. Particularly, according to Poisson effect, the compression in x and y directions comes with the increase in z direction, giving rise to the increase of dipole moments and the enhancement of internal electric field. The phenomenon is beneficial to the stable separation of photocarriers and the improvement of photocatalytic reaction efficiency.

### 3.3 Tuning optical property of $X_2$PAs monolayers by biaxial strain

In addition to the proper band structures, the ability to absorb a large amount of solar energy is also important for photocatalyst. The light absorption capacity of $X_2$PAs monolayers can be described by the light absorption coefficient $\alpha(\omega)$. $\alpha(\omega)$ is given by the following expression:

$$\alpha(\omega) = \sqrt{2}\omega[\sqrt{\varepsilon_1^2(\omega) + \varepsilon_2^2(\omega)} - \varepsilon_1(\omega)]^{1/2}, \tag{1}$$

in this function, $\varepsilon_1(\omega)$ and $\varepsilon_2(\omega)$ are the real and imaginary parts of complex dielectric functions, respectively. The $\varepsilon_2(\omega)$ is calculated as follows [37]:

$$\varepsilon_2(\omega) = \frac{4\pi^2}{m^2\omega^2} \sum_{c,v} \int_{BZ} \frac{2}{(2\pi)^3} |M_{c,v}(k)|^2 \delta(\varepsilon_{ck} - \varepsilon_{vk} - h\omega) d^3k, \tag{2}$$

where $|M_{c,v}(k)|^2$ is on behalf of the momentum operator matrix elements; $c$ and $v$ refer to conduction band and valence band, respectively. HSE06 method is applied to calculate $\varepsilon_2(\omega)$, while $\varepsilon_1(\omega)$ can be obtained from $\varepsilon_2(\omega)$ according the Kramer-Kronig relation.

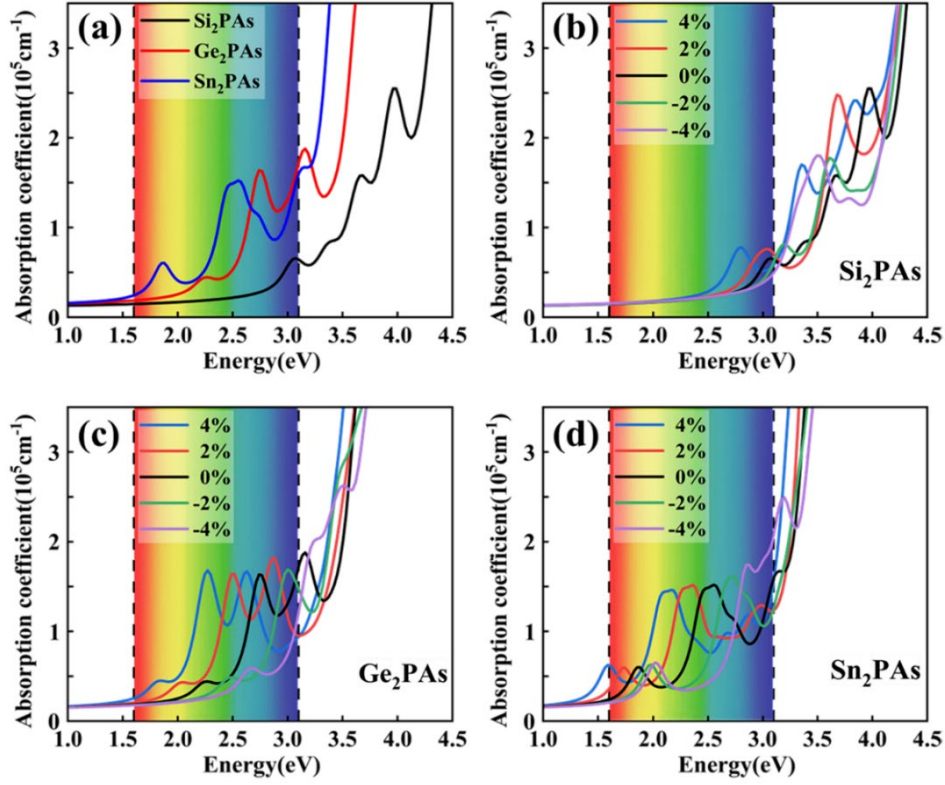

**Fig. 5.** Optical absorption coefficients of $X_2$PAs monolayers (a) without strain and (b) (c) (d) under a biaxial strain of −4% to +4%. The iridescent color is employed to mark the visible spectra (1.64~3.19 eV).

$X_2$PAs monolayers without biaxial strain are isotropic in optical properties and exhibit strong absorption capacity in the range from ultraviolet light to visible light, shown in Fig. 5(a), which will enhance their performance for photocatalytic water decomposition. In the range of visible light, Ge$_2$PAs has the maximum peak $1.64 \times 10^5$ cm$^{-1}$ at 2.74 eV, which is higher than the maximum peak $1.52\times10^5$ cm$^{-1}$ at 2.55 eV for Sn$_2$PAs by $0.12\times10^5$ cm$^{-1}$. On the contrary, Si$_2$PAs has only one absorption peak in the visible light range with a value of $0.66\times10^5$ cm$^{-1}$ with visible light photon energy of 3.08 eV. It is well known that the moderate small band gap is conducive to the absorption of visible light. Hence, Sn$_2$PAs and Ge$_2$PAs with smaller band gaps have higher visible light absorption efficiency than Si$_2$PAs. The visible light absorption coefficients of Ge$_2$PAs and Sn$_2$PAs are 2.5 times or 2.3 times higher than those of Si$_2$PAs, respectively. Here, all $X_2$PAs monolayers have good light absorption characteristics in the range of visible light energy, especially Ge$_2$PAs and Sn$_2$PAs with absorption coefficients of $10^5$ cm$^{-1}$, superior to the reported g-C$_3$N$_4$[38]. The remarkable absorption coefficients are completely comparable to perovskite minerals, indicating efficient utilization of solar energy.

The effect of strain on the light absorption characteristics of $X_2$PAs monolayers are displayed in Fig. 5 (b, c, d). As tensile strain increases, the absorption of $X_2$PAs monolayers gradually expand to the region with lower energy, showing a red shift phenomenon. On the contrary, as compressive strain increases, the absorption of $X_2$PAs gradually expands to the region with higher energy, showing a blue shift phenomenon. More interestingly, tensile strain can significantly increase the $X_2$PAs absorption efficiency within the visible light range. Therefore, applying tensile strain can enhance the absorption and utilization of visible light. These excellent properties can effectively facilitate $X_2$PAs monolayers potential applications in photocatalysis devices.

## 4. Conclusion

In generally, the electronic, optical and photocatalytic properties of $X_2$PAs ($X$ = Si, Ge and Sn) monolayers are investigated. The Janus $X_2$PAs semiconductors have the intrinsic internal electric field mainly due to the vertical asymmetry configuration, which is in favor of the stable separation of photogenerated carriers. The band edge positions and optical absorption coefficients of $X_2$PAs can be adjusted by mechanical strain to modulate the photocatalytic performance of $X_2$PAs materials. Under small strains, $X_2$PAs can match the redox potential requirement of photocatalytic water splitting. Both of the hydrogen and oxygen evolution reactions can be carried out at different sides of the $X_2$PAs monolayers. Nevertheless, under -4% compressive strains, $X_2$PAs monolayers only show high reducibility and can be used as photocatalyst for water reduction reaction to produce only hydrogen. While under +4% tensile strains, $X_2$PAs monolayers only exhibit high oxidation and can be considered as a photocatalyst for water oxidation reaction to produce only oxygen. The results indicate the Janus $X_2$PAs can be a piezo-photocatalytic switch between the OER, HER and full-reaction of redox for water splitting. With appropriate band edge position and high light absorption properties, Janus $X_2$PAs materials have a good application prospect as photocatalyst for water splitting. This study not only finds a highly efficient photocatalysts that can modulate the microscopic reaction activity through the macroscopic mechanical action, but also provides the intrinsic physical mechanism for the application of flexible piezoelectric materials in the field of photocatalysis.

**Declaration of Competing Interest**

The authors declare that they have no known competing financial interests or personal relationships that could have appeared to influence the work reported in this paper.

**Acknowledgments**

This work was supported by the National Natural Science Foundation of China (Grant No. 11474123).